\documentclass[%
 aip,
 amsmath,amssymb,
 reprint,%
]{revtex4-1}

\usepackage{graphicx}
\usepackage{dcolumn}
\usepackage{bm}
\usepackage[utf8x]{inputenc}
\usepackage[usenames,dvipsnames,svgnames,table]{xcolor}
\usepackage{selinput}
\usepackage{soul}
\usepackage[normalem]{ulem}
\usepackage{xspace}

\usepackage{layouts}

\newcommand{\MATLAB}{\textsc{Matlab}\xspace}

\begin{document}

\title{Bifurcations of front motion in passive and active Allen-Cahn-type equations}

\author{Fenna Stegemerten}
\affiliation{Institut für Theoretische Physik, Westfälische Wilhelms-Universität Münster, Wilhelm-Klemm-Str. 9, 48149 Münster, Germany}
 
\author{Svetlana Gurevich}%
\thanks{ORCID: 0000-0002-5101-4686}
\affiliation{Institut für Theoretische Physik, Westfälische Wilhelms-Universität Münster, Wilhelm-Klemm-Str. 9, 48149 Münster, Germany} 
\affiliation{Center of Nonlinear Science (CeNoS), Westfälische Wilhelms-Universität Münster, Corrensstr. 40, 48149 Münster, Germany}

\author{Uwe Thiele}
\thanks{ORCID: 0000-0001-7989-9271}
\email{u.thiele@uni-muenster.de}
\homepage{http://www.uwethiele.de}
\affiliation{Institut für Theoretische Physik, Westfälische Wilhelms-Universität Münster, Wilhelm-Klemm-Str. 9, 48149 Münster, Germany} 
\affiliation{Center of Nonlinear Science (CeNoS), Westfälische Wilhelms-Universität Münster, Corrensstr. 40, 48149 Münster, Germany}

\date{\today}

\begin{abstract}
The well-known cubic Allen-Cahn (AC) equation is a simple gradient dynamics (or variational) model for a nonconserved order parameter field. 
After revising main literature results for the occuring different types of moving fronts, we employ path continuation to determine their bifurcation diagram in dependence of the external field strength or chemical potential.
We then employ the same methodology to systematically analyse fronts for more involved AC-type models. In particular, we consider a cubic-quintic variational AC model and two different nonvariational generalisations.
We determine and compare the bifurcation diagrams of front solutions in the four considered models.
\end{abstract}

\pacs{Valid PACS appear here}
\maketitle
\textbf{The problem of front propagation has a very long history with essential contributions coming from different fields. 
One of the simplest variational model possessing front solutions is a so-called cubic Allen-Cahn (AC) equation for a nonconserved order parameter field. 
In this paper, we systematically analyse possible phase transitions in the AC equation employing analytical results given in the literature and compare them to results obtained with numerical path continuation techniques. 
Furthermore, we apply the same methodology to fronts occurring in more involved AC-type models, including a cubic-quintic variational AC model and two different nonvariational generalisations, 
where the AC equation is emended by a nonequilibrium chemical potential or is coupled to a polarisation field.}

\section{ Introduction \label{sec:level1}}
Many spatially extended systems can exist in different spatially homogeneous states that depend on various ambient parameters. In the absence of out-of-equilibrium driving forces, energy arguments hold and the homogeneous states may be (globally) stable, metastable (nonlinearly unstable) or linearly unstable. With other words they represent global minima, local minima and maxima (or saddles), respectively, of an underlying energy functional. In general, states of higher energy can be replaced through a moving front by states of lower energy. 
A simple well studied deterministic continuum model for such processes is the Allen-Cahn (AC) equation that reads in nondimensional form in one spatial dimension \cite{AlCa1979amm} [their Eq.~(11) with Eqs.~(3) and (4)],
\begin{equation}\label{eq::generalAC}
 \frac{\partial \phi}{\partial t} = - \frac{\delta \mathcal{F}}{\delta \phi}\quad\mbox{with}\quad \mathcal{F} = \int \left[ \frac{1}{2} |\partial_x\phi|^2 +f(\phi)-\mu\phi\right]\,dx.
\end{equation}
It corresponds to a nonconserved gradient dynamics on an underlying energy functional $\mathcal{F}[\phi]$ that contains a square gradient term and a local energy $f(\phi)$.
The equation arises in many different contexts and is, sometimes for specific, often quartic choices of $f(\phi)$, known as Fisher-Kolmogorov \cite{KPP1937bmm}, Fife-Greenlee \cite{FiGr1974russianmathematicalsurveys}, Schl\"ogl \cite{Schl1972zp} or Zeldovich–Frank–Kamenetsky equation \cite{ZeFr1938}. It is studied
as a model that describes dynamics in multistable systems close to phase transitions of a nonconserved order parameter field $\phi(x,t)$. 
By ``nonconserved'' we refer to a dynamics like Eq.~(\ref{eq::generalAC}) that does not conserve the total 'mass' $\int \phi(x,t) dx$ in the system. 
In its most common form, the AC equation features a double-well potential, i.e., a quartic $f(\phi)$. It may be symmetric with minima of equal energy at $\phi^-$ and $\phi^+=-\phi^-$ or be tilted by a chemical potential or external field $\mu$. 
Note that sometimes Eq.~(\ref{eq::generalAC})  with this specific $f(\phi)$ is referred to as Allen-Cahn equation \cite{MoSc2017jns}.

For example, the behavior of the magnetisation density in a ferromagnet can be described by a coarse-grained free energy density that is identical to such an energy functional. Moreover, the time evolution of this density can indeed be modeled by the AC equation \cite{Lang1992}. Although the global minimum for a symmetric double-well potential corresponds to homogeneous steady states of either $\phi^-$ or $\phi^+$, steady states consisting of arrangements of patches of $\phi^-$ and $\phi^+$ do also exist. Here, we call the interfaces between the patches ``fronts''. In the literature the notions ``kink'' and ``anti-kink'' are also used. In a fully symmetric situation the fronts are at rest. Otherwise they move, e.g., driven by $\mu\neq0$. For small driving, one can analytically determine the velocity of these fronts that connect two linearly stable states corresponding to local minima of $f(\phi)$, see e.g.\ Refs.~\onlinecite{Pismen2006b,Misbah2016}.

Via another type of moving front, a linearly stable state invades a linearly unstable state \cite{Saar2003prspl}. Such fronts occur when systems are suddenly quenched into an initially homogeneous unstable state. Small local perturbations then grow and develop into patches of stable states that spread out over the whole domain. In general, the stable state may correspond to a pattern. This case can, for example, be found in Taylor-Couette flow \cite{AhCa1983prl}. Fronts also appear in Rayleigh-B\'enard systems when the heat flux is suddenly increased. Then a convective vortex front propagates into the unstable conductive state \cite{FiSt1987prl}. Moreover, front propagation into unstable states is studied in crystal growth \cite{Lang1980rmp} and in the context of chemical reactions \cite{Burger1985}. 
Another distinction is between pulled and pushed fronts, where the velocity of the former is controlled by linear effects while in the latter case nonlinear effects dominate \cite{Pismen2006b,Misbah2016}.
Here, we investigate the dependence of the motion of fronts on the strength of the external field or chemical potential $\mu\neq0$. 
On the one hand, we consider the well-known case of a simple AC system with a double well potential \cite{Pismen2006b}, and use it to introduce our methodology that is based on numerical continuation \cite{EGUW2019springer}. 
On the other hand, we consider AC models with more complicated local energy \cite{Lajz1981f,BeLT1991prl} that allows for a larger number of front types and an active AC model that can not be written as a gradient dynamics, i.e., in the variational form~(\ref{eq::generalAC}).
We employ two ways to render the AC model nonvariational. 
First, we add a nonvariational term analogously to Ref.~\onlinecite{ACGW2017pre}, i.e., we incorporate a nonequilibrium chemical potential \cite{EGUW2019springer} as also frequently done for mass-conserving Cahn-Hilliard-type dynamics \cite{WTSA2014nc,CaTa2015arcmp,STAM2013prl,RaBZ2019epje}. 
Such an amended AC equation may be employed to model, e.g., front propagation in a liquid crystal light valve \cite{ACGW2017pre}. Second, we couple the AC equation for the order parameter to an evolution equation of a polarisation field $P$ in a similar spirit as in an active phase-field-crystal (PFC) model \cite{MeLo2013prl,OpGT2018pre,ELWG2012ap}. For other models including an active Swift-Hohenberg equation see Ref.~\onlinecite{KoTl2007c}, a tentative systematics in the case of single scalar fields is given in the introduction of Ref.~\onlinecite{EGUW2019springer}. 
Taking the second option for an active AC model in one spatial dimension results in a system which is similar to the FitzHugh-Nagumo model \cite{Fitz1961biophysicaljournal,NaAY1962proceedingsoftheire} describing spike generation in squid axons. There, the evolution equation of the membrane potential corresponds to an AC equation with cubic nonlinearity.
As the AC model with double-well local energy is arguably the simplest nonlinear gradient dynamics model for a parity symmetric spatially extended system with $\phi\to-\phi$ symmetry, its active generalisations are likely the simplest such active models. Their detailed understanding will help to extend our knowledge of the collective behaviour of active systems consisting of a large number of active particles which are able to transform different types of energy into motion.
Here, we focus on front motion, however, the range of phenomena in such systems is much richer.
In general, the microscale constituents interact in such a way that on a macroscopic (collective) level directed collective motion and clustering phenomena may occur \cite{ZBFS2010pnasu}. Different forms of interaction result in different phenomena, e.g., a purely repulsive interaction may give rise to motility-induced phase separation \cite{CaTa2015arcmp}, whereas a combination of repulsive and attractive interaction allows for the formation of swarms, e.g., of fish or birds \cite{Sumpter2010}, bacteria colonies \cite{PSJS2012prl} or cell motion \cite{SSGJ2007pre, SPCG2013pcb}.
The occurring collective structures may consist of disordered or well ordered arrangements that are referred to as active clusters and active crystals \cite{PSSP2013s}, respectively.
Our work is structured as follows: First, we focus in section~\ref{sec::PS} on the passive systems, i.e., AC systems that evolve towards equilibrium. In particular, section~\ref{sec::AC} introduces the cubic AC equation, briefly reviews linear and nonlinear marginal stability analyses, and determines different front types and their velocities as a function of driving strength $\mu$. In section~\ref{sec::pqAC}, we consider the quintic AC equation that has already attracted much interest, e.g., in Refs.~\onlinecite{Lajz1981f,BeLT1991prl, Saar1989pra, CoCh1997pd, PoNJ1991pra}. In contrast to the literature, we focus on the transitions occuring in the behaviour of fronts when the driving strength $\mu$ is changed.
Second, section~\ref{sec:aAC} considers the two nonvariational amendments of the AC equation. Thereby, section~\ref{sec:acAC} considers the first option, namely, the case of a nonequilibrium chemical potential. We discuss the analytical expression for the shift in front velocity close to the transition from the variational to the nonvariational case and compare this to our fully nonlinear numerical results. Section~\ref{sec:aACP} considers the second option namely the coupling of a cubic  AC equation with the dynamics of a polarisation field. It is shown that a condition for onset of motion can be determined analytically, similar to the case of the active PFC model in Ref.~\onlinecite{OpGT2018pre}. 
Bifurcation diagrams are presented for both cases of active AC equations that summarise the behaviour of all occurring front solutions when the driving strength is varied. Finally, in section~\ref{sec:conc} we conclude and give an outlook to future work.

 \section{Passive Systems}\label{sec::PS}
%
\subsection{General form of Allen-Cahn-type models}\label{sec::AC}
%
We start our analysis with the well-known standard passive version of the AC equation that describes the time evolution of an one-dimensional nonconserved order parameter field $\phi=\phi(\tilde{x},t)$. The general form is obtained by introducing the energy $\mathcal{F}$ in Eq.~(\ref{eq::generalAC}) into the general gradient dynamics. It reads
 \begin{equation}\label{eq::generalpassAC}
 \frac{\partial \phi}{\partial t}=\frac{\partial^{2} \phi}{\partial \tilde{x}^{2}}-f^{\prime}(\phi)+\mu\, ,
 \end{equation}
 where $\mu$ is the chemical potential, $f^{\prime}(\phi)$ is the first derivative of the purely nonlinear local energy density $f$ with respect to $\phi$, and $\tilde{x}$ is the coordinate in the laboratory frame. In order to find the velocity $v$ of steadily travelling fronts propagating between two homogeneous steady states $\phi^\mathrm{s}$ given by $-f^{\prime}(\phi^\mathrm{s})+\mu=0$, 
 we transform the system~\eqref{eq::generalpassAC} into a co-moving frame $x=\tilde{x}-vt$ and get
 \begin{equation}\label{eq::comovinggenAC}
-v  \frac{\textrm{d} \phi}{\textrm{d} x}=\frac{\textrm{d}^{2} \phi}{\textrm{d} x^{2}}-f^{\prime}(\phi)+\mu\, .
 \end{equation}
Multiplying Eq.~\eqref{eq::comovinggenAC} by $\dfrac{\textrm{d}\phi}{\textrm{d}x}$ and integrating (employing Neumann boundary conditions) yields the explicit expression for the velocity
\begin{equation}
v=\frac{f\left(\phi^\mathrm{s1}\right)-f\left(\phi^\mathrm{s2}\right)+\mu\left(\phi_{s2}-\phi_{s1}\right)}{\int_{-\infty}^{\infty}\left(\frac{\textrm{d}\phi}{\textrm{d}x}\right)^{2}\textrm{d}x}\, ,
\end{equation}
for a front propagating from $\phi^\mathrm{s2}$ at $x\to-\infty$ into $\phi^\mathrm{s1}$ at $x\to\infty$.

In the following, we consider two specific passive systems in sections~\ref{sec::cAC} and \ref{sec::pqAC} below, namely, AC equations with cubic and cubic-quintic nonlinearities, respectively. By employing $\mu$ as main control parameter we focus on the influence of a physically most relevant quantity, that can be easily controlled externally. Note, that this differs from most employed and well-known parametrisations \cite{BBDK1985pd,PoNJ1991pra,EbSa2000pd,ArWe1978aim,PCGO1994prl}. 
%

\begin{figure*}
  \centering 
   \includegraphics[width=1\textwidth]{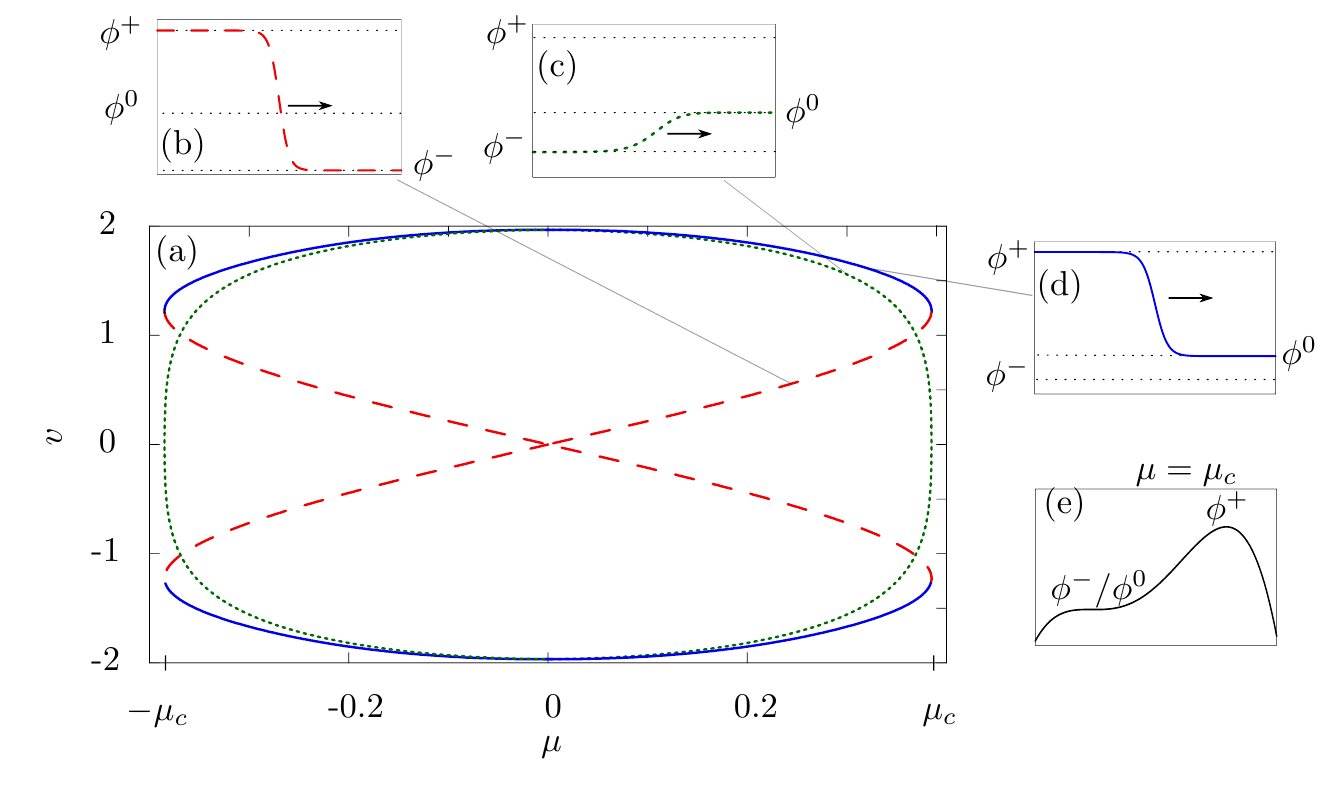}
   \caption{The central panel (a) presents the bifurcation diagram of front states described by the cubic AC equation \eqref{eq::pAC} in terms of the front velocity $v$ as a function of the driving strength $\mu$. The red dashed lines correspond to fronts between the two linearly stable homogeneous states $\phi^{-}$ and $\phi^{+}$ with an example given in panel~(b). 
   The two respective fronts corresponding to the two different linearly stable states invading the unstable states are indicated by green dotted and blue solid lines with examples in panels (c) and (d), respectively. 
   Panel (e) gives the equivalent mechanical potential $V(\phi)$ at the critical value $\mu=\mu_{c}$ where $\phi^{-}$ and $\phi^{0}$ annihilate.}\label{fig:bif-ac}
 \end{figure*}
 
\subsection{The passive cubic Allen-Cahn equation}\label{sec::cAC}
%
\subsubsection{Model}
%
In the case of the cubic Allen-Cahn equation, $f(\phi)=-\frac{1}{2}\phi^{2}+\frac{1}{4}\phi^{4}$ and hence, in the co-moving frame we obtain:
\begin{equation}\label{eq::pAC}
-v\frac{\textrm{d} \phi}{\textrm{d} x}=\frac{\textrm{d}^{2}\phi}{\textrm{d}x^{2}}+\phi-\phi^{3}+\mu \,.
\end{equation}
As mentioned above, this equation is extensively studied in the literature. Here, we review some main approaches to introduce the methodology for our analysis of the more involved models, c.f.~e.g., the textbooks Refs.~\onlinecite{Pismen2006b,Misbah2016}. For the cubic nonlinearity, three homogeneous steady state solutions $\phi^{0}(\mu)$, $\phi^{+}(\mu)$ and $\phi^{-}(\mu)$ exist for $-\mu_{c}<\mu<\mu_{c}$ with $\mu_{c}=\frac{2}{3\sqrt{3}}\approx 0.385$. Here, $\phi^{+}$ and $\phi^{-}$ are linearly stable states while $\phi^{0}$ is unstable. 
Equation~\eqref{eq::pAC}  can be seen as a mechanical system, where $\phi$ corresponds to the position and $x$ is time. In this framework, Eq.~(\ref{eq::pAC}) describes a particle moving in a double well potential $V(\phi)=-f(\phi)+\mu\phi$ with friction $v$. As for $\mu=0$ the maxima of $V(\phi)$ are of equal height, we find that for any friction $v>0$ the particle will always asymptotically approach the 'position' $\phi^{0}=0$, which corresponds to the unstable homogeneous state. 
However, in reality one observes, that such fronts moving into the unstable state $\phi^{0}$, have the certain specific velocity \cite{ArWe1978aim}, namely, $v=2$. Hence, a dynamical selection of the front speed occurs (for details see, e.g.,~Ref.~\onlinecite{Saar1988pra}), that we briefly discuss next.
%

\subsubsection{Linear marginal stability analysis}\label{sec:lmsa}
%
The selection problem is tackled employing a linear marginal stability analysis: We consider the linearisation of the fully time-dependent AC equation about the unstable homogeneous state $\phi^{0}$, i.e., we consider the leading edge of the front (cf.~Ref.~\onlinecite{Saar1988pra}). Note that at the  marginal stability point, i.e., the point where the resulting eigenvalues are zero, the group velocity of perturbations at the leading edge of the front,
\begin{equation}
v_{g}=\frac{\textrm{d}\omega^{r}(k)}{\textrm{d} k^{r}}
\end{equation}
equals the velocity of the front
 \begin{equation}
 v_{f}=\frac{\omega^{r}(k)}{k^{r}}.
 \end{equation}
Here, $k$ and $\omega$ are the wavenumber and frequency, respectively, and the superscript $r$ denotes the real part. In this way one obtains a linear marginal velocity
\begin{equation}\label{eq::linmag}
 v_{l}:=v_{f}=v_{g}=\pm2\sqrt{-f^{\prime\prime}(\phi^{0})}\, .
 \end{equation}
 Thus, at the linear marginal stability point perturbations can not grow above the moving front profile and any front with $v<v_{l}$ is unstable. Hence, Eq.~\eqref{eq::linmag} provides a criterion to determine the dynamically selected front velocity. In Ref.~\onlinecite{Saar1988pra} it is shown that the linear marginal velocity is an attractive fixed point, such that any front moving with $v>v_{l}$ eventually converges to one  with $v=v_{l}$. This is referred to as a pulled front.

\subsubsection{Numerical path continuation}
%
Next, we determine fronts numerically and compare them to the analytical results. In particular, we employ path continuation techniques \cite{Kuznetsov2010,DWCD2014ccp} bundled in the \MATLAB toolbox \texttt{pde2path}~\cite{UeWR2014nmma,EGUW2019springer} or in the continuation package \texttt{auto07p} \cite{DoKK1991ijbc,DFSC2007}
to determine front profiles and velocities in dependence of the parameter $\mu$ that acts as the strength of a driving caused by an external field or chemical potential \footnote{ If not stated otherwise we employed \texttt{pde2path}}.

The resulting bifurcation diagram is shown in terms of the front velocity in Fig.~\ref{fig:bif-ac}~(a). Interestingly, three branches exist corresponding to three different front types that are illustrated by the examples in panels (b) to (d). 
The overall symmetry of the bifurcation diagram reflects the symmetries $(x,v)\to(-x,-v)$ and $(\phi,\mu)\to (-\phi,-\mu)$ of Eq.~(\ref{eq::pAC}).
The three front types are a front between the two linearly stable homogeneous states $\phi^{-}$ and $\phi^{+}$ indicated by the red dashed lines, and the two respective fronts corresponding to the two different linearly stable states invading the unstable state indicated by green dotted and blue solid lines. 
The velocity $v$ of the front between stable states is determined as explained in section~\ref{sec::AC}. For the front that moves to the right at positive $\mu$ (red dashed line in the upper right quadrant of Fig.~\ref{fig:bif-ac}~(a)), $v$ first linearly than faster increases with increasing $\mu$. 
The curve becomes vertical when $\mu$ reaches the critical value $\mu_{c}$ where the lower maximum of the mechanical potential ($\phi^{-}$) and the minimum ($\phi^{0}$) annihilate (see Fig.~\ref{fig:bif-ac}~(e)). This implies that the bifurcation curve passes a saddle-node bifurcation and folds back.
Then it continues towards lower $\mu$ as solid blue line representing fronts where the globally stable state (linearly stable and of lowest energy) invades the unstable state $\phi^{0}$. To understand their origin, in the next section we introduce the nonlinear marginal analysis.
Decreasing $\mu$ further, when crossing $\mu=0$ the fronts where the globally stable state $\phi^{+}$ invades the unstable state $\phi^{0}$ become fronts where the metastable state $\phi^{-}$ (linearly stable, but of higher energy than the globally stable state) invades $\phi^{0}$ (green dotted lines).

Note, finally, that at the value of $\mu$ where the red dashed and green dotted line cross, fronts from $\phi^{0}$ to $\phi^{-}$ and from $\phi^{-}$ to $\phi^{+}$ have the same velocity.

  \begin{figure}
  \centering 
\includegraphics[width=0.5\textwidth]{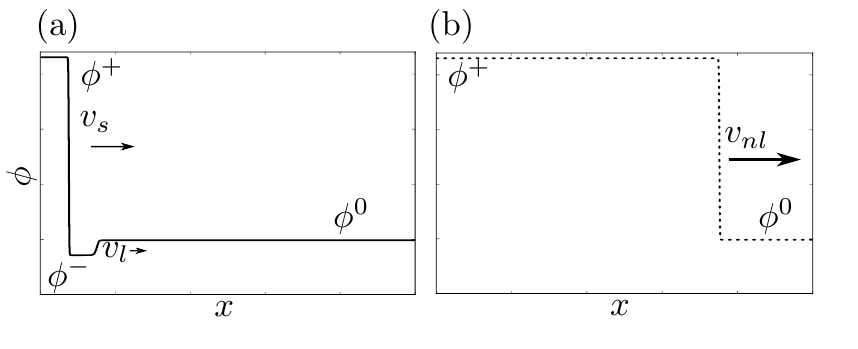}
\caption{Snapshots from a time simulation of Eq.~\eqref{eq::pAC} at $\mu=0.377$ for a front initially composed of a front between the two stable states $\phi^{+}$ and $\phi^{-}$ moving at $v_s$ and the linear marginal front between $\phi^{-}$ and $\phi^{0}$ moving at $v_l<v_s$ [see (a)]. At a later time the two fronts merge into a single nonlinear marginal front between $\phi^{+}$ and $\phi^{0}$ moving at $v_{nl}>v_s$ [see (b)].
}\label{fig:coupl-front}
\end{figure}

  \begin{figure}
  \centering 
\includegraphics[width=0.5\textwidth]{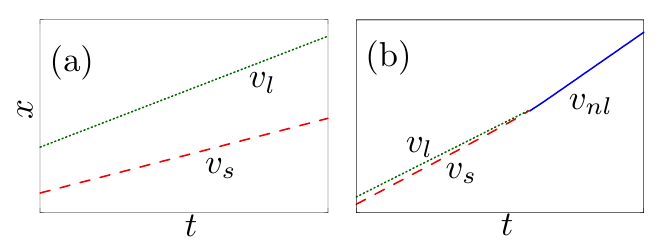}
\caption{The motion of the fronts illustrated in Fig.~\ref{fig:coupl-front} is illustrated in space-time plots of the front position (a)  at $\mu=0.350$ where $v_l>v_s$, i.e., the fronts do not merge, and (b) at $\mu=0.375$ where $v_l<v_s$, i.e., the fronts eventually merge and move with  $v_{nl}$.}\label{fig:sp-tim}
\end{figure}
This implies that in the vicinity of this point one may expect moving structures consisting of two fronts as depicted exemplarily in Fig.~\ref{fig:coupl-front}~(a). To the left of this point the  $\phi^{-}$ to $\phi^{+}$ front stays behind the $\phi^{0}$ to $\phi^{-}$ front (cf.~Fig.~\ref{fig:sp-tim}~(a)) while to the right of this point the former catches up with the latter. Then they merge as indicated in Fig.~\ref{fig:sp-tim}~(b)  and create the faster $\phi^{0}$ to $\phi^{+}$ front shown in Fig.~\ref{fig:coupl-front}~(b).
%
\subsubsection{Nonlinear marginal stability analysis}
%
Using the mechanical analogy we can also understand the second occurring front moving into the unstable state: A particle moving in this potential may start at either of the maxima corresponding to $\phi^{+}$ or $\phi^{-}$. In Ref.~\onlinecite{Saar1989pra, Saar1988pra} it is shown, 
that solving the fully nonlinear equation above a threshold $\mu$ one additionally finds a so-called invasion front, where $\phi^{+}$ invades $\phi^{0}$, that is marginally stable and has a nonlinear velocity $v_{nl}>v_{l}$.

  \begin{figure}
  \centering 
\includegraphics[width=0.5\textwidth]{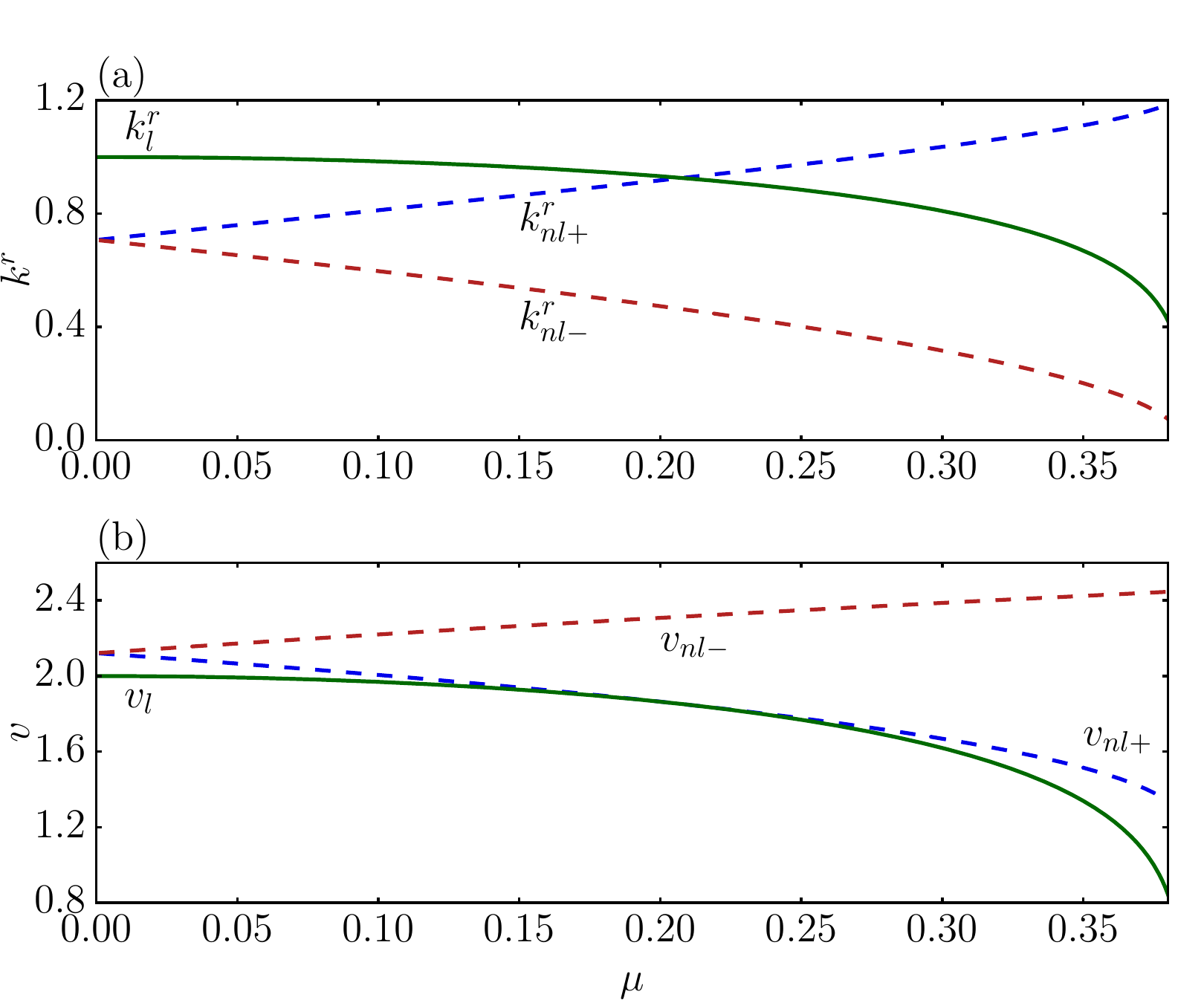}
   \caption{
Shown are the (a) real part of wave numbers $\textrm{Re}(k)=k^r$ and (b) velocities $v$ of linear (subscript ``l'') and nonlinear (subscript ``nl'') marginal fronts as a function on driving strength $\mu$. 
The nonlinear marginal results are given for fronts where state $\phi^{+}$ invades state $\phi^{0}$ (subscript ``$+$'') as well as for fronts where state $\phi^{-}$ invades state $\phi^{0}$ (subscript ``$-$''). 
The implication of the dependencies for the selection of front velocities is discussed in the main text.
}\label{fig:nonlinear-exp}
\end{figure}

Therefore, we study \eqref{eq::pAC} to find the front solution corresponding to the blue solid line in Fig.~\ref{fig:bif-ac}~(a) following the ideas given in Refs.~\onlinecite{BBDK1985pd, Saar1988pra,Saar1989pra,Pismen2006b}.
We denote the two possible linearly stable states by $\phi^{\pm}$. Because the front is monotonic we can define
\begin{equation}\label{eq:ansatz-h}
h(\phi)=\frac{\textrm{d}\phi}{\textrm{d} x}
\end{equation}
as a function of $\phi$.
As the front connects two homogeneous states we request
\begin{equation}\label{eq:neumann}
h\left(\phi^{0}\right)=h\left(\phi^{\pm}\right)=0\, .
\end{equation}
Deriving $\eqref{eq:ansatz-h}$ with respect to $x$ yields
\begin{equation}\label{eq:h-ddx}
\frac{\textrm{d}\phi}{\textrm{d} x}\frac{\textrm{d}h}{\textrm{d} \phi}=h(\phi)\frac{\textrm{d}h}{\textrm{d} \phi}=\frac{\textrm{d}^{2}\phi}{\textrm{d} x^{2}}\, .
\end{equation}
Thus, inserting \eqref{eq:ansatz-h} and $\eqref{eq:h-ddx}$ into \eqref{eq::pAC} we get:
\begin{equation}\label{eq:h-dif}
-vh(\phi)=h(\phi)\frac{\textrm{d}h}{\textrm{d} \phi}+\phi-\phi^{3}+\mu\, .
\end{equation}
Finally, with a power series ansatz up to second order for $h(\phi)$ we obtain
\begin{equation}\label{eq:v-nl}
v_{nl}(\mu)=\frac{3}{\sqrt{2}}\left(\phi^{0}+\phi^{\pm}\right)
\end{equation}
for the nonlinear marginal velocity. Note, that considering an ansatz of higher order, the comparison of coefficients always results in a second order polynomial. Therefore, the obtained expression is an exact solution to the cubic AC equation.

Figure~$\ref{fig:nonlinear-exp}$~(a) shows the linear marginal wave number $k^{r}_{l}$ (solid green line, obtained in section~\ref{sec:lmsa}) and the here obtained nonlinear marginal wave numbers $k^{r}_{nl+}$ and $k^{r}_{nl-}$ (dashed blue and red line, respectively) as a function of $\mu$, 
whereas Fig.~$\ref{fig:nonlinear-exp}$~(b) gives the corresponding velocities $v_{l}$, $v_{nl+}$, and $v_{nl-}$. Subscripts ``$+$'' and ``$-$'' refer to fronts where states $\phi^{+}$ and $\phi^{-}$ invade state $\phi^{0}$, respectively. 
The dynamical selection always implies that the steeper (larger wave number) and slower front is chosen. Considering the front connecting $\phi^{-}$ and $\phi^{0}$, we see that $k^{r}_{l}>k^{r}_{nl-}$ for all $\mu$, i.e., at all $\mu$ where the front exists, the linear marginal velocity is selected (solid line in Fig.~\ref{fig:nonlinear-exp}~(b),
corresponding to the numerically obtained green dotted line in Fig.~\ref{fig:bif-ac}).
In contrast, for the  front connecting $\phi^{+}$ and $\phi^{0}$, the condition $k^{r}_{l} >k^{r}_{nl+}$ only holds for $\mu<0.2078$, while at larger $\mu$ one has $k^{r}_{nl+}>k^{r}_{l}$. 
This implies that for $\mu<0.2078$ we find the linear marginal velocity (solid line in Fig.~\ref{fig:nonlinear-exp}~(b)) while for $\mu>0.2078$ the nonlinear marginal velocity (dashed blue line in Fig.~\ref{fig:nonlinear-exp}~(b)) is selected. This perfectly agrees with the numerical results in Fig.~\ref{fig:bif-ac} (solid blue line).

In summary, for $\mu<0.2078$ the two fronts are degenerate as both move with  the linear marginal velocity, while at larger $\mu$ two distinct velocities are found, i.e., above $\mu=0.2078$ the dotted green and the solid blue branch in Fig.~\ref{fig:bif-ac}~(a) differ, while they exactly coincide below this value. 
Note that the $v(\mu)$ curves in Fig.~\ref{fig:nonlinear-exp} are tangential to each other where the two corresponding $k^{r}(\mu)$ curves cross. This implies that a clear transition in the corresponding numerically obtained curve can be best seen when inspecting the second derivative of $v$ with respect to $\mu$.

Note, that all described front types are analytically studied in Ref.~\onlinecite{Pismen2006b}. We have seen that the numerical results obtained by path continuation well agree with the analytical results. The given brief review and comparison provides the starting point for our analysis of more complex models.
%
\subsection{The passive cubic-quintic Allen-Cahn equation}\label{sec::pqAC}
\begin{figure*}
  \centering 
  \includegraphics[width=0.95\textwidth]{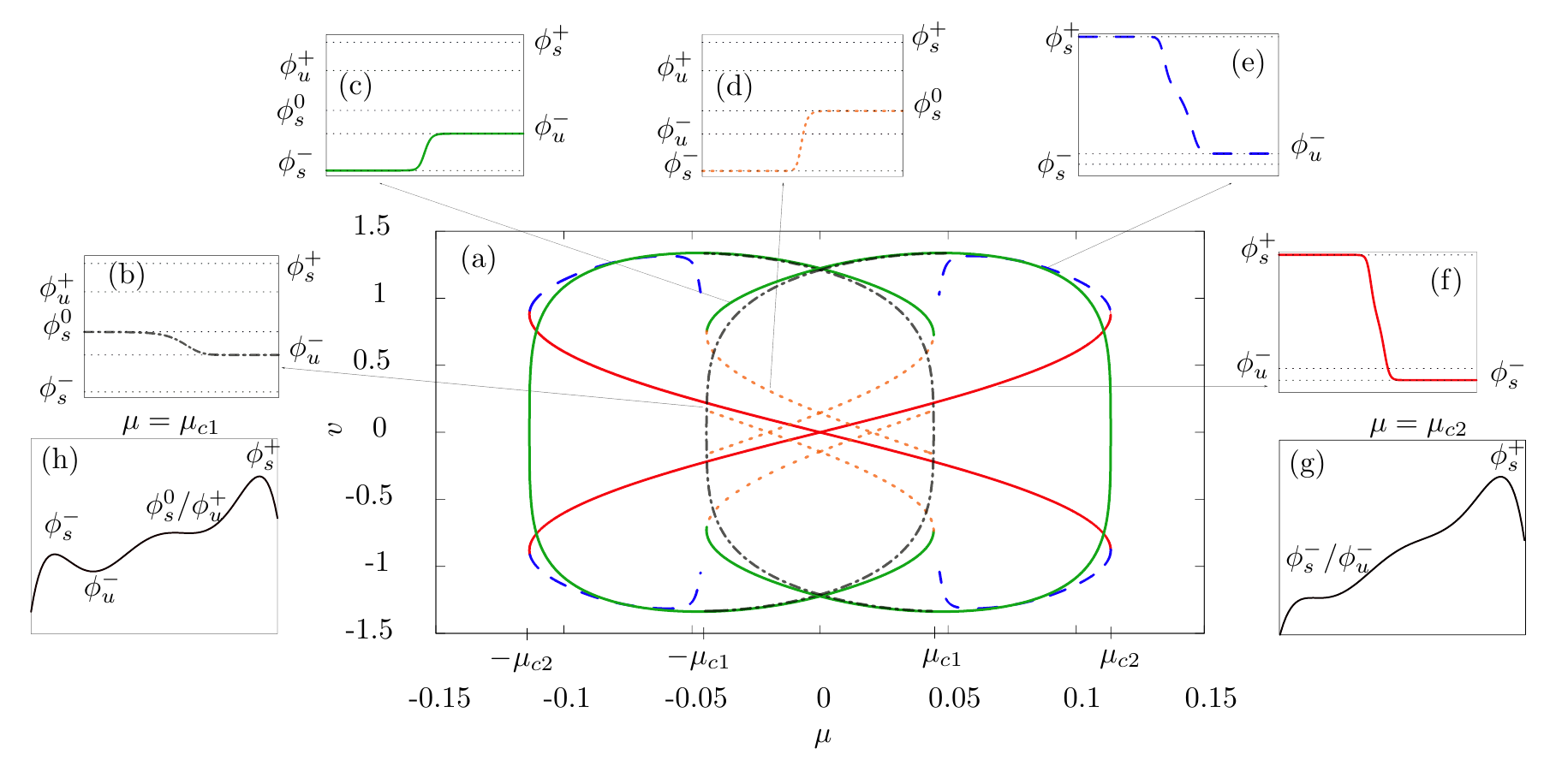}
   \caption{
The central panel (a) presents the bifurcation diagram of front states described by the cubic-quintic AC equation \eqref{eq::pqAC} in terms of the front velocity $v$ as a function of the driving strength $\mu$. 
The red solid lines correspond to fronts between the two linearly stable states $\phi_{s}^{+}$ and $\phi_{s}^{-}$ with an example given in panel~(f) for $\mu=0.09$. At the saddle-node bifurcation at $\mu=\mu_{c2}$ the states $\phi_{s}^{-}$ and $\phi_{u}^{-}$ annihilate. 
The green solid lines corresponds to fronts between $\phi_{s}^{-}$ and $\phi_{u}^{-}$ (panel~(c) for $\mu=-0.02$) or between $\phi_{s}^{+}$ and $\phi_{u}^{+}$ depending on the sign of $\mu$. 
At the saddle-node bifurcation at $\mu=\mu_{c1}$ the states $\phi_{s}^{0}$ and $\phi_{u}^{+}$ annihilate (at $\mu=-\mu_{c1}$ the states $\phi_{s}^{0}$ and $\phi_{u}^{-}$).
The orange dotted lines refer to front solutions between $\phi_{s}^{+}$ and $\phi_{s}^{0}$ or between $\phi_{s}^{-}$ and $\phi_{s}^{0}$ (panel~(d) for $\mu=-0.02$), again depending on the sign of $\mu$. 
The blue dashed lines refer to front solutions between $\phi_{s}^{+}$ and $\phi_{u}^{-}$ (panel~(e) for $\mu=0.09$). Finally, the grey dot-dashed lines represent front solutions between $\phi_{s}^{0}$ and $\phi_{u}^{+}$ or between $\phi_{s}^{0}$ and $\phi_{u}^{-}$ (panel~(b) for $\mu=-0.02$). 
The thin horizontal black dotted lines in front profile panels represent the homogeneous steady states, i.e., the real roots of $f^{\prime}(\phi)-\mu$ in \eqref{eq::pqAC}. Panels (g) and (h) gives the equivalent mechanical potential $-f(\phi)+\mu\phi$ at the critical values $\mu_{c1}$ and $\mu_{c2}$, respectively.}\label{fig:bif-qac}
 \end{figure*}

%
Next, we analyse the cubic-quintic AC equation, i.e., Eq.~(\ref{eq::generalAC}) with ($f(\phi)=\frac{1}{6}\phi^{6}-\frac{a}{4}\phi^{4}+\frac{b}{2}\phi^2$) similar to the one studied in Ref.~\onlinecite{BeLT1991prl} in the context of the creation of metastable phases in crystallisation processes. It is also studied in Ref.~\onlinecite{Kueh2015sjuq} with the additional influence of stochastic noise.
The equation for steadily moving fronts (\ref{eq::comovinggenAC}) is then
\begin{equation}\label{eq::pqAC}
-v\frac{\textrm{d} \phi}{\textrm{d} x}=\frac{\textrm{d}^{2}\phi}{\textrm{d}x^{2}}-\phi^{5}+a\phi^{3}-b\phi+\mu \, ,
\end{equation}
It has stable homogeneous solutions $\phi_{s}^{-}$, $\phi_{s}^{0}$, $\phi_{s}^{+}$ and unstable homogeneous solutions $\phi_{u}^{-}$ and $\phi_{u}^{+}$, all depending on $\mu$. As in Ref.~\onlinecite{BeLT1991prl} we use $a=5/4$ and $b=1/4$.
The corresponding bifurcation diagram
is depicted in Fig.~\ref{fig:bif-qac}~(a) together with exemplary front profiles in panels (b)-(f). Panels (g) and (h) show the potential in the mechanical analogy at the critical values $\mu_{c1}\approx0.044$ and $\mu_{c2}\approx0.114$, where two homogeneous steady states annihilate.
To understand the rich picture involving a number of different branches of front solutions, we first identify the two linear marginal velocities corresponding to fronts invading the two unstable states. We also characterise all nonlinear marginal velocities. 
Note that the overall symmetries of the bifurcation diagram correspond to the ones known from Fig.~\ref{fig:bif-ac}.

\begin{figure}
  \centering 
  \includegraphics[width=0.5\textwidth]{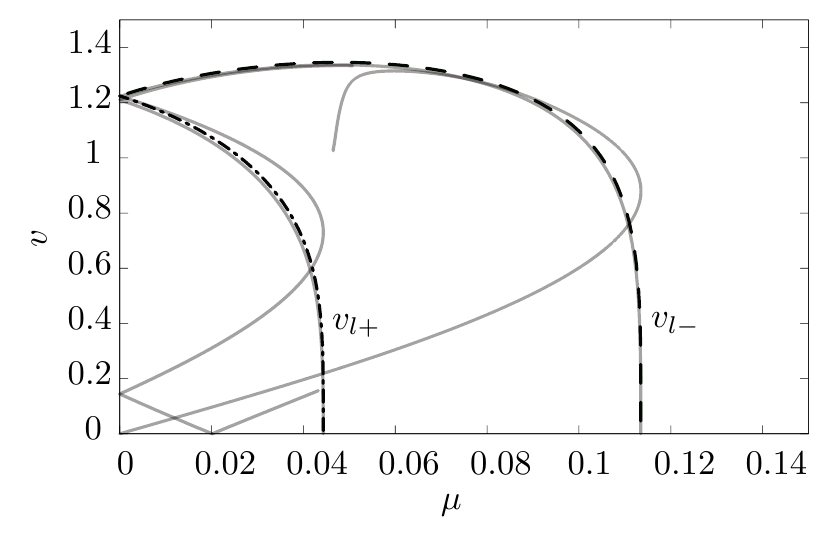}
 \caption{The linear marginal velocity $v_{l-}$ found in Eq.~\eqref{eq::pqAC} corresponds to the black dashed line whereas $v_{l+}$ corresponds to the black dot-dashed line. The results gained with $\texttt{auto-07p}$ are illustrated as the grey solid lines.}\label{fig:v-l-bech}
 \end{figure}

Because the nonlinear function in \eqref{eq::pqAC} is a quintic polynomial, the roots as a function of $\mu$ can not be calculated analytically. 
Nevertheless, we can find them numerically, e.g., employing continuation. In consequence, we are able to study the linear marginal velocity by inserting the numerical results into the specific form of Eq.~(\ref{eq::linmag}), i.e., into
\begin{align}
v_{l+}&=\pm2\sqrt{-5{\phi_{u}^{+}}^{4}+3a{\phi_{u}^{+}}^{2}-b}\\
v_{l-}&=\pm2\sqrt{-5{\phi_{u}^{-}}^{4}+3a{\phi_{u}^{-}}^{2}-b}\, ,
\end{align}
where $v_{l+}$ and $v_{l-}$ are the linear marginal velocities for a front invading the unstable states $\phi_{u}^{+}$ and $\phi_{u}^{-}$, respectively.
They are depicted in Fig.~\ref{fig:v-l-bech} as black dot-dashed and dashed lines, respectively. The numerical results are indicated by the grey solid lines and in particular for the linear marginal velocities well coincide with the semi-analytical result.

In this way we identify the grey dot-dashed branches and the green solid branches for $\lvert\mu\rvert>\mu_{c1}$ in Fig.~\ref{fig:bif-qac}(a) as linear marginal, i.e., pulled fronts. 
We note that different parts of the structure of the bifurcation diagram for the simple cubic AC equation presented above in Fig.~\ref{fig:bif-ac}~(a) can be recognised as substructures within the bifurcation diagram for the quintic case in Fig.~\ref{fig:bif-qac}~(a).
For instance, the red solid branches corresponding to fronts between the two stable states $\phi_{s}^{+}$ and $\phi_{s}^{-}$ in Fig.~\ref{fig:bif-qac}(a) behave similarly to the red dashed branches in Fig.~\ref{fig:bif-ac}~(a) over the entire $\mu$-range. 
The blue dashed and green solid branches in Fig.~\ref{fig:bif-qac}(a) are similar to blue solid and green dotted branches in Fig.~\ref{fig:bif-qac}~(a), however, only for $\lvert\mu\rvert>\mu_{c1}$.

Therefore, we expect the blue dashed branch in Fig.~\ref{fig:bif-qac}~(a) to correspond to a nonlinear marginal front, as they can be identified as invasion fronts at the bifurcation point, where it merges with the red branch. 
Moreover, in the interval $-\mu_{c1}<\mu<\mu_{c1}$ we twice observe a structure similar to the one of the cubic AC equation, once for negative and once for positive velocities. Again, this can be explained referring to the mechanical analogy: 
In this $\mu$-range we can consider the potential as being composed of two cubic AC potentials. At the critical values $\lvert\mu\rvert=\mu_{c1}$ the metastable state merges with one of the unstable states and thus for $\lvert\mu\rvert>\mu_{c1}$ there is only one 'cubic' AC potential left.

We now return to the blue dashed front in Fig.~\ref{fig:bif-qac}~(a), for a profile see Fig.~\ref{fig:bif-qac}~(e). To understand why the front does not exist for $\lvert\mu\rvert<\mu_{c1}$ we study again the mechanical analogy at $\mu=\mu_{c1}$ illustrated in Fig.~\ref{fig:bif-qac}~(h): Consider a particle starting at $\phi_{s}^{+}$ and just reaching $\phi_{u}^{-}$ without overshooting. Let the corresponding friction be $v_{1}$. Considering now a particle that moves from $\phi_{s}^{+}$ to $\phi_{s}^{0}$, the corresponding friction $v_{2}$ needs to satisfy $v_{2}>v_{1}$. 
Moreover, a particle starting in $\phi_{u}^{-}$  requires more negative friction to move up to $\phi_{s}^{+}$ than it does to move up to $\phi_{s}^{0}$. Hence, we claim $v_{1}>v_{3}$, where $v_{3}$ is the friction to move from $\phi_{s}^{0}$ to $\phi_{u}^{-}$. 
That is, we require $v_{3}<v_{1}<v_{2}$, where $v_{3}$ corresponds to the velocity $v_{3}\approx1.33$ of a front on the grey dot-dashed branch at $\mu_{c1}$. However, $v_{2}$ is the velocity at the bifurcation point, where the orange dotted branch becomes the green solid branch with $v_{2}\approx0.73<v_{3}$.
Hence, $v_{1}$, the velocity of the blue dashed solution does not satisfy the condition at $\mu=\mu_{c1}$ and therefore, the branch can not exist anymore. Note, that this argument is similar to one given in Ref.~\onlinecite{BeLT1991prl} and we can identify the blue dashed front solutions as being similar to those found there, however, as our equations slightly differs they are not identical.
The question how exactly this branch ends is not easy to answer. Our calculations show that following the branch towards smaller $\mu$, the slight shoulder visible in Fig.~\ref{fig:bif-qac}~(e) develops into a long inclined plateau of increasing length. This indicates that even at $\mu>\mu_{c1}$ the front already ``feels'' the presence of the two additional homogeneous steady states that exist for $\mu\le\mu_{c1}$. With other words, the spatial dynamics slows down close to these ``ghost solutions'' forming the sloped plateau. The slope gets smaller the longer the plateau becomes, i.e., the closer one approaches $\mu_{c1}$ from above. We are not able to follow the blue dashed branch further than shown in Fig. ~\ref{fig:bif-qac}~(a) as the plateau becomes too long for the largest of our numerical domain sizes. Our hypothesis is that the curve becomes vertical when approaching $\mu_{c1}$ from above where the structured front between $\phi_{s}^{+}$ and $\phi_{u}^{-}$ decays into several fronts between $\phi_{s}^{+}$, $\phi_{u}^{+}$, $\phi_{s}^{0}$ and $\phi_{u}^{-}$ that exist for $\mu\le\mu_{c1}$. 


%
\section{Active systems}\label{sec:aAC}
So far, we have examined front motion described by variational AC equations, i.e., systems that can be written as nonconserved gradient dynamics. They describe systems that ultimately approach thermodynamic equilibrium. 
Next, we consider active systems where this is not the case. In particular we consider two different examples of nonvariational extensions of the cubic AC equation.
%
\subsection{Nonvariational cubic Allen-Cahn equation}\label{sec:acAC}
In the first considered example of a nonvariational AC equation a term $g_{nv}$ is added to \eqref{eq::generalpassAC} that breaks the gradient dynamics form (\ref{eq::generalpassAC}). The resulting equation is \cite{ACGW2017pre}
\begin{align}
\frac{\partial\phi}{\partial t}&=\frac{\partial^{2} \phi}{\partial \tilde{x}^{2}}+\phi-\phi^{3}+\mu+\epsilon g_{nv}\left(\phi,\frac{\partial\phi}{\partial  \tilde{x}},\frac{\partial^{2}\phi}{\partial  \tilde{x}^{2}}\right)\notag\\
\textrm{with}\qquad &g_{nv}\left(\phi,\frac{\partial\phi}{\partial  \tilde{x}},\frac{\partial^{2}\phi}{\partial  \tilde{x}^{2}}\right)=\left(\frac{\partial \phi}{\partial  \tilde{x}}\right)^{2}+2b\phi\frac{\partial^{2} \phi}{\partial  \tilde{x}^{2}}\label{eq::gnv}\, .
\end{align}
Note, that the complete equation becomes variational for $b=1$. 

\begin{figure}
  \centering 
  \includegraphics[width=0.5\textwidth]{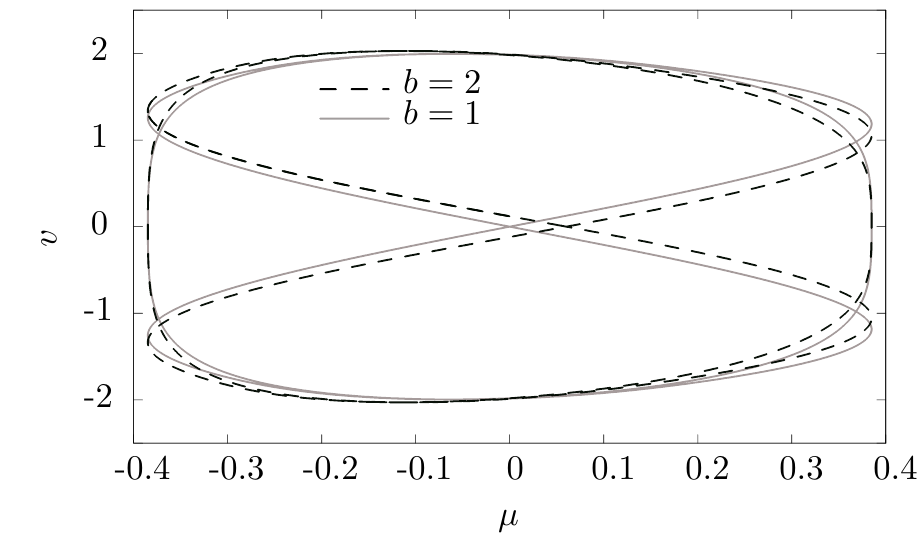}
    \caption{Comparison of bifurcation diagrams for the variational cubic AC equation with $b=1$ (grey solid) and the nonvariational cubic AC equation~\eqref{eq::nonv-v} with $b=2$ (black dashed). 
    The shift in velocity $v_{1}$ for a front propagating from $\phi^{+}$ into $\phi^{-}$ at $\mu=0$ for $\epsilon=0.2$ is consistent with the analytical result.}\label{fig:nonv-2}
 \end{figure}

Again, we are interested in the velocities of fronts and their dependence on the external driving strength $\mu$. 
Following Ref.~\onlinecite{ACGW2017pre}, we use a front solutions of the variational Eq.~\eqref{eq::pAC} (i.e., Eq.~\eqref{eq::gnv} with $\epsilon=0$) as a reference solution $\phi_{F}$, and employ the ansatz $\phi( \tilde{x},t)=\phi_{F}( \tilde{x}-vt) + u(\tilde{x}-vt,\epsilon t)$. 
Here, $v\approx v_{0}-\epsilon v_{1}$ with $v_{0}$ being the velocity of the reference front, i.e., $\phi_{F}( \tilde{x}-v_{0}t):=\phi_{F}(x)$.

One multiplies Eq.~\ref{eq::gnv} by $\partial\phi/\partial  \tilde{x}$, integrates in $x$, employing the point symmetry $\phi_{F}(x)=-\phi_{F}(-x)$ and expands in $\epsilon$ (see the Appendix for details).
Then, to linear order one can write the nonvariational contribution as \cite{ACGW2017pre}
\begin{equation}\label{eq::nonv-v}
v_{1}=\frac{\int_{-\infty}^{\infty}g_{nv}\left(\phi_{F},\frac{\partial\phi_{F}}{\partial x},\frac{\partial^{2}\phi_{F}}{\partial x^{2}}\right)\partial_{x}\phi_{F}\, \textrm{d}x}{\int_{-\infty}^{\infty}\left(\partial_{x}\phi_{F}\right)^{2}\,\textrm{d}x}\, .
\end{equation}
We notice that to this order $v_{1}$, in contrast to $v_{0}$ does not depend on the difference in energy densities $f$, but only on the nonvariational part $g_{nv}$ and the shape of the reference front shape $\phi_{F}(\mu)$.
Figure~\ref{fig:nonv-2} gives the corresponding numerically obtained bifurcation diagram of front solutions to Eq.~\eqref{eq::nonv-v} for a nonvariational case with $b=2$ and relatively small $\epsilon=0.2$ (black dashed line) in comparison to the variational case with $b=1$ (grey solid line). 
Due to $g_{nv}$, the symmetry $\mu\rightarrow -\mu$ is broken because $g_{nv}$ does not have the $\phi\rightarrow -\phi$ symmetry.

Inserting $\phi_{F}$  into \eqref{eq::nonv-v} and solving the integrals for $\mu=0$ where $v_{0}=0$ yields $v_{1}=\frac{2\sqrt{2}}{5}$, i.e., in the nonvariational case the front moves even without external driving. 
The shift in velocity for the chosen $\epsilon$ coincides on the scale of Fig.~\ref{fig:nonv-2} with the numerical result. 
At larger strength $\epsilon$ of the nonvariational term, linear considerations do not suffice anymore (not shown). Note that the point of zero velocity for the fronts connecting $\phi^-$ and $\phi^+$ is with increasing $\epsilon$ shifted towards positive $\mu$. It may be seen as the out-of-equilibrium coexistence point of the two states connected by the front.

\subsection{Cubic Allen-Cahn equation coupled to polarisation field}\label{sec:aACP}
%
As second and final example of a nonvariational AC equation we consider a cubic AC equation coupled in a simple way to the linear dynamics of a polarisation field. 
Here we focus on the one-dimensional case where $P$ describes the local strength of directional order and employ the same coupling between order parameter field $\phi$ and $P$ as employed in the active phase-field-crystal (PFC) model \cite{MeLo2013prl,MeOL2014pre,ChGT2016el,OpGT2018pre}.
The system of equations then reads
\begin{equation}\label{eq:aAC}
\begin{aligned}	\frac{\partial \phi}{\partial t}&=-\frac{\delta \mathcal{F}}{\delta \phi}-\alpha_{0}\frac{\partial P}{\partial x}\,,\\
	\frac{\partial P}{\partial t}&=D_{T}\frac{\partial^{2}}{\partial x^{2}}\frac{\delta \mathcal{F}}{\delta P} - D_{r}\frac{\delta \mathcal{F}}{\delta P}-\alpha_{0}\frac{\partial \phi}{\partial x}\,,\\
	\mathrm{with}\qquad \mathcal{F}[\phi,P]&=\mathcal{F}_{AC}[\phi]+\mathcal{F}_{P}[P]
\end{aligned}
\end{equation}
where
\begin{equation}
\mathcal{F}_{AC}=\int \,\textrm{d}x\,\left[\frac{1}{2}\left(\frac{\partial}{\partial x}\phi\right)^{2}-\frac{1}{2}\phi^{2}+\frac{1}{4}\phi^{4}-\mu\phi\right]
\end{equation}
is the energy for the cubic passive AC equation as used before, 
\begin{equation}
\mathcal{F}_{P}=\int \,\textrm{d}x\,\,\frac{1}{2} P^{2}
\end{equation}
is the energy of the polarisation $P$ that favours a state of random orientation ($P=0$) and does not allow for spontaneous polarisation. Both coupling terms have a strength $\alpha_0$ (called ``activity'') and their form represents the simplest possible form to couple the scalar $\phi$ and the 'vector' $P$. Furthermore, $D_{T}$, and $D_{r}$ are positive translational and rotational diffusivities, respectively. Note that the coupling terms break the gradient dynamics structure.

 \begin{figure}
  \centering 
  \includegraphics[width=0.5\textwidth]{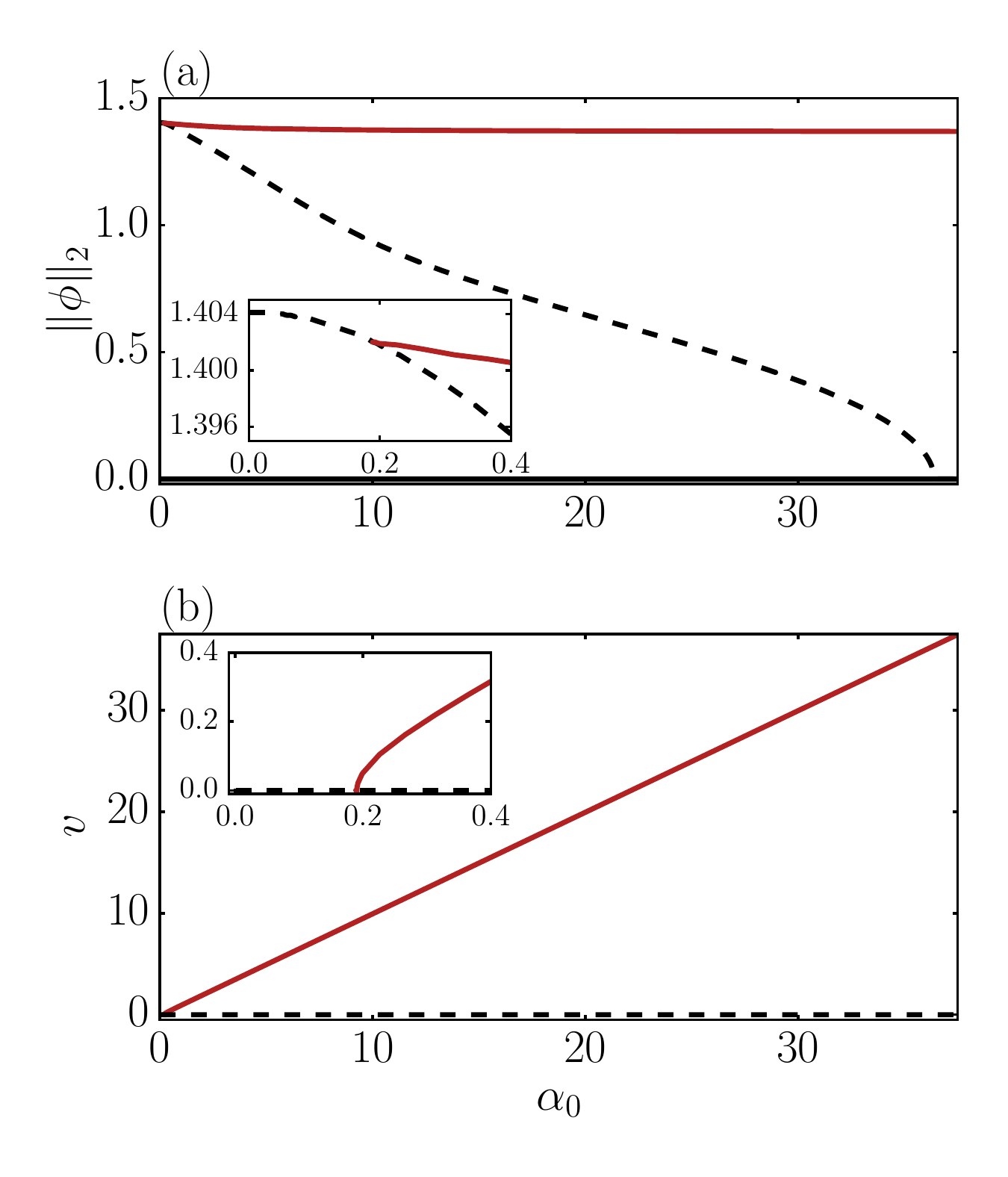}
  \caption{Bifurcation diagram for resting and moving fronts for the active cubic AC equation \eqref{eq:aAC} at $\mu=0$. 
  Panels (a) and (b) show the norm $||\phi||_{2}$ and the front velocity $v$ in dependence of activity $\alpha_{0}$, respectively. 
  The resting and moving fronts are given as black dashed and red solid lines, respectively. The insets focus on the region where the drift pitchfork bifurcation occurs. The black solid line represents the homogenous solution branch.}
  \label{fig:bif-aac-l2}
   \end{figure}

In analogy to the active PFC model we expect that steady states of the passsive model ($\alpha_0$) remain at rest until they undergo a drift pitchfork bifurcation at a critical activity where they start to move. We determine an analytical criterion for the onset of motion in analogy to the derivation for the active PFC model in Ref.~\onlinecite{OpGT2018pre}. It reads
\begin{equation}\label{eq::onset-of-motion}
 0=\Vert\partial_{x}\phi_{s}\Vert^{2}-\Vert\partial_{x} P_{s}\Vert^{2}\, ,
 \end{equation}
 and allows us to determine the critical values of the activity parameter $\alpha_{0}$. Here, $\phi_{s}$ and $P_{s}$ denote steady state profiles whose $L_2$-norm is taken. Note, that the specific criterion differs from the one in Ref.~\onlinecite{OpGT2018pre} as there the order parameter field $\phi$ follows a conserved dynamics while here the dynamics is nonconserved. 
 This is further discussed in Ref.~\onlinecite{Ophaus2019Muenster}.

 Fig.~\ref{fig:bif-aac-l2} presents results of continuation runs following front solutions of Eqs.~(\ref{eq:aAC}) at $\mu=0$ using the activity $\alpha_0$ as control parameter. The black dashed line corresponds to resting fronts connecting $\phi^-$ and $\phi^+$. These resting fronts become shallower with increasing $\alpha_0$ and the branch terminates on the branch of trivial homogeneous states at about $\alpha_0=36.3$. 
 However, already at rather small $\alpha_{0}\approx 0.19$, a drift pitchfork bifurcation occurs (see inset of Fig.~\ref{fig:bif-aac-l2}~(a)) where a branch of steadily moving fronts (red solid line) emerges from the branch of resting fronts. The numerical results confirm the criterion (\ref{eq::onset-of-motion}). 
 In the limit of large activity we find $v=\alpha_{0}$, as depicted in Fig.~\ref{fig:bif-aac-l2}~(b). This agrees with analogous results in Ref.~\onlinecite{OpGT2018pre} where also an analytical argument is given.
 
 \begin{figure*}
  \centering 
  \includegraphics[width=1.\textwidth]{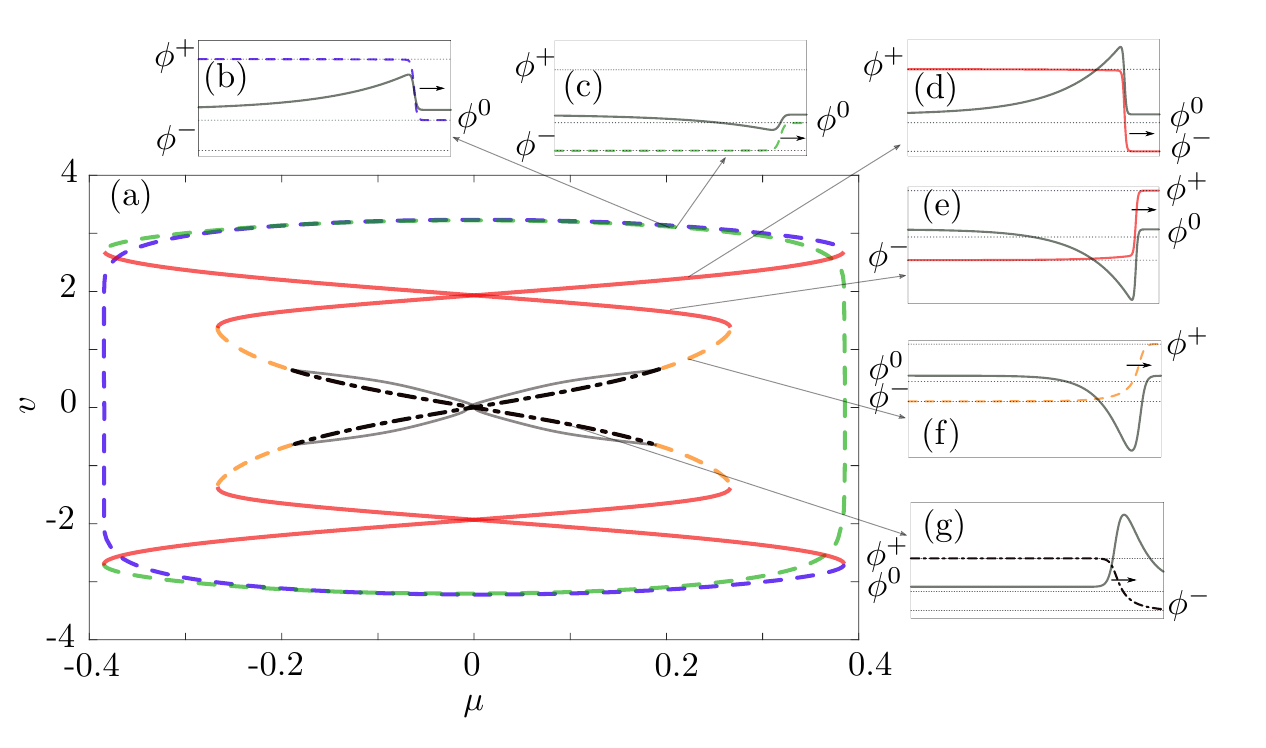}
  \caption{(a) Bifurcation diagram of front states described by the active Allen-Cahn model~\eqref{eq:aAC}. Shown is the front velocity as a function of the external driving strength $\mu$ at fixed activity $\alpha_0=1.9951$.
    There are various fronts between  the two stable states $\phi^-$ and $\phi^+$, namely, the red solid, black dot-dashed, grey solid and orange dashed lines. The blue and green dashed lines correspond to fronts propagating into the unstable state $\phi^{0}$. 
The grey solid line corresponds to a branch of time periodic oscillating fronts.
Panels (b) to (g) show selected front profiles $\phi(x)$ in colors equal to the corresponding branch in panel~(a) at $\mu\approx 0.2$. The accompanying polarisation profiles $P(x)$ are given in grey. For completeness, we also indicate the steady homogeneous states $\phi^-$, $\phi^0$ and $\phi^+$ as dotted horizontal lines. Note however, that the solution panels are differently scaled on the $\phi,P$ axes as the polarisation profiles differ in magnitudes.
}\label{fig:bif-aac}
  \end{figure*}

Next, we take a moving front at a particular value of activity ($\alpha_0=1.9951$) from Fig.~\ref{fig:bif-aac-l2} and perform a continuation in $\mu$ to obtain a bifurcation diagram in analogy to Fig.~\ref{fig:bif-ac} for the passive cubic AC equation. 
The result is presented in Fig.~\ref{fig:bif-aac}~(a). Panels (b) to (g) show selected front profiles $\phi(x)$ in colors equal to the corresponding branch in panel~(a). 
The accompanying polarisation profiles $P(x)$ are given in grey. For completeness, we also indicate the steady homogeneous states $\phi^-$, $\phi^0$ and $\phi^+$ as dotted horizontal lines. Note that the front location is off center in the computational domain as the relaxation to the homogeneous states can be very asymmetric.
 

The fronts on the red solid and black dot-dashed branches at $\mu=0$ correspond to the moving and resting front in Fig.~\ref{fig:bif-aac-l2}. Increasing $\mu$, the symmetry of the potential in $\phi$ is broken and two red branches emerge. 
Both of them undergo saddle-node bifurcations at respective critical values $0.26$ and $\mu_c$. The larger one is identical to $\mu_c$ in section~\ref{sec::cAC} as the homogeneous steady states do not depend on activity (in contrast to their stability). 
The smaller one can not be obtained from an analysis of the local equilibrium potential $f(\phi)-\mu$. 

\begin{figure}
  \centering 
  \includegraphics[width=0.5\textwidth]{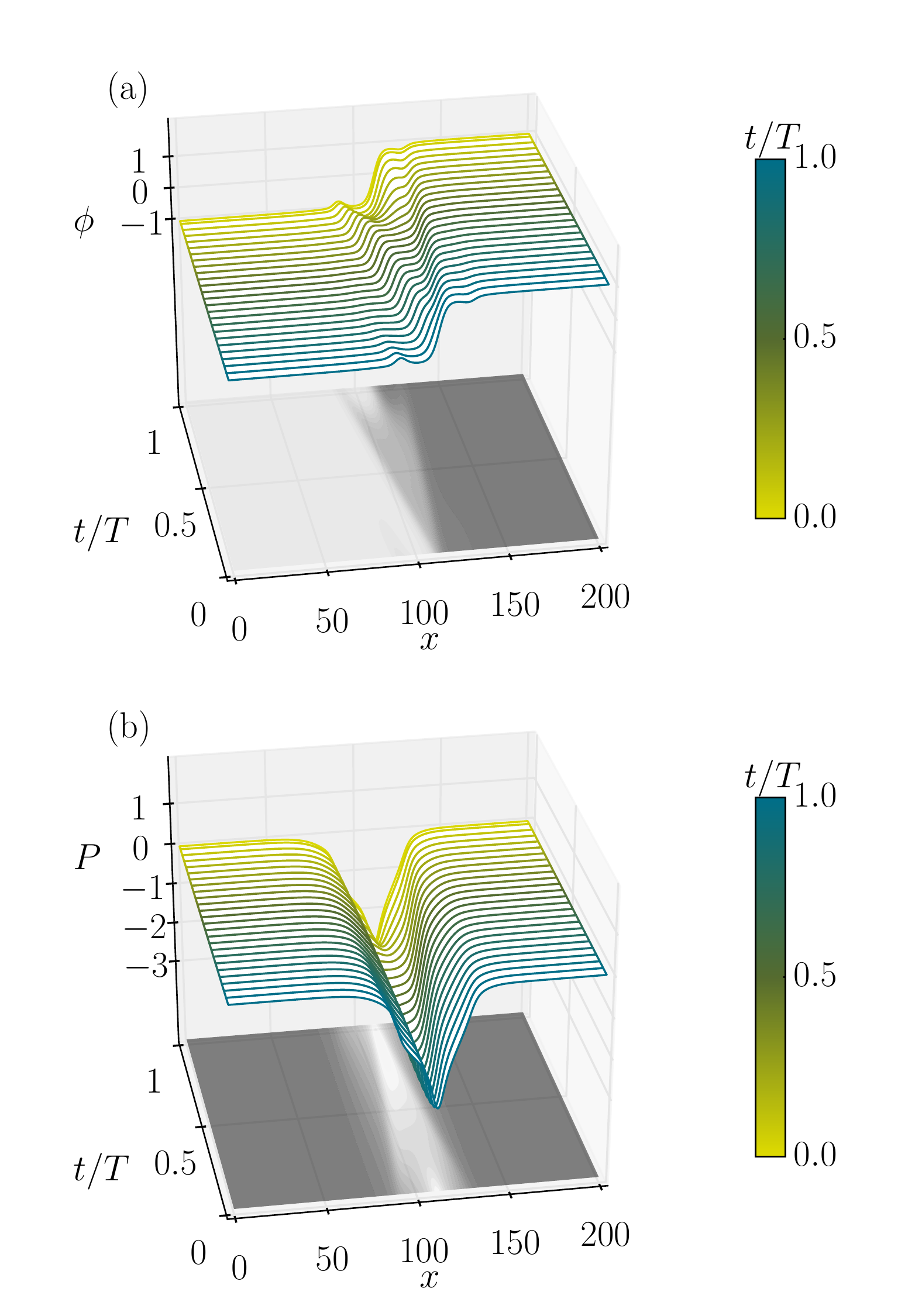}
  \caption{Space-time plots for a periodically modulated front solution of Eq.~\eqref{eq:aAC}. Shown are (top) the concentration profile and (bottom) the polarisation profile at $\mu=0.2$ in the frame moving with their mean velocity $v=0.15$. The different colors of the contour indicate time scaled by the period $T\approx14.5$. Furthermore, a projection of the profiles is shown on the $x-t$ plane in grey-scale.}\label{Fig:hopf}
  \end{figure}

Another interesting feature are fronts that do not move steadily but are modulated in a time-periodic manner. They are found on the grey solid branches in Fig.~\ref{fig:bif-aac}~(a) and emerge with finite oscillation period of about $T=12.5$ from a Hopf bifurcation where orange dashed and black dot-dashed branch connect. The branch continues towards smaller $\mu$ with a monotonically increasing period (up to about $T=14.5$ at $\mu=0$).
Typical changes in the moving solution profiles over one period of time are depicted in Fig.~\ref{Fig:hopf}.

Overall one can state that the coupling of a simple AC dynamics to a linear dynamics of a polarisation field introduces a number of additional unstable front solutions all connecting $\phi^{+}$ and $\phi^{-}$. Hence, the structure of the bifurcation diagram strongly changes from the passive case in Fig.~\ref{fig:bif-ac}~(a) and the active case in Fig.~\ref{fig:bif-aac}~(a). Note that due to the coupling, also the marginal stable fronts propagating into the unstable state become unstable in the active model. 
%
 \section{Conclusion}\label{sec:conc}
%
 In the present work we have investigated front solutions in a number of passive and active Allen-Cahn (AC) equations employing continuation techniques. We have focused on the dependency of front velocities on an external driving $\mu$, e.g., a chemical potential or external field. The results have been presented in the form of bifurcation diagrams. First, we have reviewed the widely available analytical results for the simple cubic AC equation and have introduced the concepts of pulled and pushed fronts employing linear and nonlinear marginal stability analysis. We have highlighted that there exist fronts that change their character from pulled to pushed as the driving $\mu$ is increased across a threshold.

Next we have extended the analysis to the cubic-quintic AC equation that allows for more homogeneous steady states and, in consequence, for more fronts connecting them. We have presented a rather involved bifurcation diagram employing again the driving $\mu$ as control parameter. It shows the various front solutions connecting the up to five homogeneous steady states. 
In general, our results allow one to better understand how the different fronts are related and how they transform with increasing driving strength. To understand substructures of the diagram it has been helpful to discuss symmetries and how parts of the  cubic-quintic potential resemble the cubic potential. 

The considered model is similar to the one studied in Ref.~\onlinecite{BeLT1991prl} in the context of the creation of metastable phases in crystallisation processes. In particular, they investigate how double-fronts emerge that first create a metastable phase before it is transformed into the stable phase. 
Their main control parameter is the order parameter value corresponding to one of the thermodynamic phases, while here we have kept the energy functional fixed and employed the chemical potential (external field) as control parameter. 
We have shown how fronts and transitions similar to the ones discussed in  Ref.~\onlinecite{BeLT1991prl} are embedded into the full bifurcation diagram of front states.

As Ref.~\onlinecite{TuBe1992pra} extends the discussion of Ref.~\onlinecite{BeLT1991prl} to systems where two order parameter fields are coupled, in the future it may be interesting to revisit also such and more complicated two- or multi-field models employing continuation techniques.

After considering the cubic and cubic-quintic AC equations that represent gradient dynamics models, we have considered nonvariational extensions. First, we have analysed a cubic AC equation with the addition of a standard nonvariational chemical potential (see, e.g., classification in the introduction of Ref.~\onlinecite{EGUW2019springer}). We have found that main symmetries of the bifurcation diagram are broken, however, occurring front types and overall topology of the diagram remain the same as in the passive case. Our numerical results agree at small strength of the nonvariational influence $\epsilon$ with the approximate analytical results of Ref.~\onlinecite{ACGW2017pre} but show some deviations at larger $\epsilon$. Here it will be interesting in the future to employ continuation techniques to investigate fronts in the nonvariational cubic-quintic AC equation recently introduced in Ref.~\onlinecite{ACCG2019oe} as a model for a liquid crystal light valve experiment with optical feedback.

Changes with respect to the passive case are more dramatic in the second nonvariational model that couples a cubic AC equation with a linear equation of a polarisation field, similar to such couplings in more involved models of active matter. In this case we have encountered fronts that move due to activity even at zero external driving. They emerge in a drift pitchfork bifurcation similar to the onset of motion in active phase-field-crystal \cite{OpGT2018pre} and active Cahn-Hilliard models \cite{Ophaus2019Muenster}. This then implies a much richer bifurcation diagram that even contains oscillating front states that emerge at a Hopf bifurcation of steady fronts. We believe future comparative studies that analyse front motion and its emergence in a larger class of systems would be highly valuable.


\appendix
\section{Nonvariational cubic Allen-Cahn equation}

As in the literature \cite{ACGW2017pre} the derivation of Eq.~\ref{eq::nonv-v} is only sketched, and we find it instructive, here we reproduce it in greater detail. 
The aim is to determine an analytical expression for the velocity of moving fronts that are solutions of the nonvariational AC equation (\ref{eq::gnv}) and to discuss its dependence on the strength of the nonvariational influence. 

We introduce the ansatz
\begin{align}\label{eq::ansatz-ACGW2017pre}
\phi( \tilde{x},t)&=\phi_{F}( \tilde{x}-vt)+u( \tilde{x}-vt,\epsilon t)\notag\\
v&=v_{0}-\epsilon v_{1}
\end{align}
where $\phi_{F}$ is the solution to \eqref{eq::pAC} with velocity $v_{0}$, i.e., $\phi_{F}( \tilde{x}-v_{0}t):=\phi_{F}(x)$ as the velocity changes due to the nonvariational term. Moreover, we add a small adjustment function $u$ also of order $\epsilon$. 
For the ansatz \eqref{eq::ansatz-ACGW2017pre} we introduce $\xi= \tilde{x}-vt=x+\epsilon v_{1}t$. Linearizing $\phi_{F}(\xi)$ around $x$ yields
\begin{equation}\label{eq::linphif}
\phi_{F}(\xi)=\phi_{F}(x)+\epsilon v_{1}\frac{\partial\phi_{F}}{\partial\xi}\bigg|_{\xi=x}+\mathcal{O}(\epsilon^{2})
\end{equation}
Inserting the ansatz $\eqref{eq::ansatz-ACGW2017pre}$ into $\eqref{eq::gnv}$ using $\eqref{eq::linphif}$ and linearizing in $\epsilon$ yields

\begin{align}\label{eq::ACGW2017pre-3}
-&\epsilon v_{1} v_{0}\frac{\partial^{2}\phi_{F}}{\partial \xi^{2}}\bigg|_{\xi=\xi}+\epsilon v_{1}\frac{\partial\phi_{F}}{\partial \xi}\bigg|_{\xi=\xi}-v_{0}\frac{\partial u}{\partial \xi}\notag\\
=&\epsilon v_{1}\frac{\partial^{3} \phi_{F}}{\partial \xi^{3}}\bigg|_{\xi=\xi}
+\epsilon v_{1}\frac{\partial\phi_{F}}{\partial \xi}\bigg|_{\xi=\xi}-3\epsilon v_{1}\phi_{F}^{2}\frac{\partial\phi_{F}}{\partial \xi}\bigg|_{\xi=\xi}\notag\\
&+\frac{\partial^{2} u}{\partial \xi^{2}}+ u -3\phi_{F}(\xi)^{2} u+\epsilon g_{nv}\left(\phi_{F}(\xi),\frac{\partial\phi_{F}}{\partial \xi},\frac{\partial^{2}\phi_{F}}{\partial \xi^{2}}\right)\bigg|_{\xi=\xi}\, .
\end{align}
Deriving \eqref{eq::pAC} with respect to $x$ 
\begin{equation}
-v_{0}\frac{\partial}{\partial x} \frac{\partial\phi_{F}(\xi)}{\partial x}=\frac{\partial}{\partial x}\frac{\partial^{2} \phi_{F}(x)}{\partial x^{2}}+\frac{\partial \phi_{F}(x)}{\partial x}-3\phi_{F}(x)^{2}\frac{\partial \phi_{F}(x)}{\partial x} \, ,
\end{equation}
and introducing the linear operator
\begin{equation}\label{eq::lino}
L=-v_{0}\frac{\partial}{\partial x}-\frac{\partial^{2}}{\partial x^{2}}-1+3\phi_{F}(x)^{2}\, ,
\end{equation}
we obtain 
\begin{equation}\label{eq::transl}
L^{\dagger}\left( \frac{\partial}{\partial x} \phi_{F}(-x)\right)=0\, ,
\end{equation}
where $L^{\dagger}$ is the adjoint of $L$.
Next, we identify the linear operator $\eqref{eq::lino}$ in \eqref{eq::ACGW2017pre-3} 
\begin{equation}\label{eq::ACGW2017pre-zw}
\epsilon v_{1} \frac{\partial\phi_{F}}{\partial \xi}\bigg|_{\xi=x}+\epsilon v_{1}L\left( \frac{\partial}{\partial x} \phi_{F}\right)+Lu
=\epsilon g_{nv}\bigg|_{\xi=x}\, .
\end{equation}
With \eqref{eq::transl}, \eqref{eq::ACGW2017pre-zw} simplifies to
\begin{equation}\label{eq::adjust}
L u= \epsilon g_{nv}\left(\phi_{F},\frac{\partial\phi_{F}}{\partial x},\frac{\partial^{2}\phi_{F}}{\partial x^{2}}\right)- \epsilon v_{1}\frac{\partial\phi_{F}}{\partial x}
\end{equation}
According to the Fredholm alternative \cite{Ramm2001tamm} \eqref{eq::adjust} is only solvable if 
\begin{equation}\label{eq::scalar}
\left.\left<\epsilon g_{nv}\left(\phi_{F},\frac{\partial\phi_{F}}{\partial x},\frac{\partial^{2}\phi_{F}}{\partial x^{2}}\right)- \epsilon v_{1}\frac{\partial\phi_{F}}{\partial x}\right|  \frac{\partial}{\partial x} \phi_{F}\right>=0\, ,
\end{equation}
because \eqref{eq::transl} has a nontrivial solution.

\end{document}